\documentclass[12pt]{article}
\usepackage{amssymb,amsmath,amsthm,amsfonts,amscd}
\usepackage{graphicx}
\newcommand\be{\begin{equation}}
\newcommand\ee{\end{equation}}

\newcommand{\fatalpha}{{\bf \alpha \kern -0.44em \alpha}}
\newcommand{\fatsigma}{{\bf \sigma \kern -0.54em \sigma}}
\newcommand{\tpchi}{{\bf \chi \kern -0.35em \chi}}
\newcommand{\llambda}{{\bf \lambda \kern -0.45em \lambda}}

\bibliography{plain}
 \title{\bf  One-Mode Wigner Quasi-probability Distribution Function for Entangled Coherent States Generated by Beam Splitter and Cavity QED}
 \vspace{20mm}
\author{G. Najarbashi \thanks{E-mail: Najarbashi@uma.ac.ir} ,
  S. Mirzaei \thanks{E-mail: SMirzaei@uma.ac.ir}
\\
{\small Department of Physics, University of Mohaghegh Ardabili, P.O. Box 179, Ardabil, Iran.} \\
 \\
}\pagebreak
\begin{document}
\maketitle \vspace{0mm}

\maketitle \vspace{0mm}
\begin{abstract}
 In this paper, we use the displacement operator together with parity operation to construct the superposition of two coherent states. By transmitting this superposition from 50-50 beam splitter the two-mode qubit like ECS is generated. Moreover, we introduce a controllable method for producing qutrit like ECS using atom-field interaction in cavity QED and beam splitter. We will show that the distances of peaks of Wigner functions for reduced density matrices of  two-mode  ECS's are entanglement sensitive and can be a witness for entanglement. To confirm the results we use concurrence measure to compare bipartite entanglement of ECS's with the behavior of peaks of Wigner functions. Moreover, we investigate decoherence effects on Wigner function, arising from transmitting ECS's through noisy channels.

  {\bf PACs Index:03.65.Ud}
\end{abstract}

\section{Introduction}
Phase-space representations of quantum states have been important tools for exploring the connections between
quantum and classical physics. In 1932, Wigner introduced a distribution function in mechanics that permitted a description of mechanical phenomena in a phase space \cite{wigner1,wigner2}. Some important characteristics of the spatial Wigner function of entangled photon pairs were analyzed in \cite{Torres}. In \cite{Banerji} the negativity of the Wigner function was discussed as a measure of the non-classicality. Negativity of the Wigner function is the reason why the Wigner function can not be regarded as a real probability distribution but a quasi-probability distribution function and it is a good indication of the possibility of the occurrence of nonclassical properties. Wigner functions have been especially used for describing the quadratures of the electrical field with coherent and squeezed states or single photon states \cite{Breitenbach,Lvovsky,Agarwal1}. The four-dimensional chronocyclic Wigner function of the parametric downconversion state was discussed in \cite{Benjamin}.

In 1926, Schrodinger introduced coherent states \cite{Schrodinger} then advanced studies were done in \cite{Glauber,Perelomov}. In recent years, there has been the considerable interest in studying multi-mode quantum states of radiation fields because they have widely role in quantum information theory. The problem of generating various quantum states of an electromagnetic field was discussed in \cite{Enk2,Sanders3,Milburn1,Milburn2,Tanas1,Tanas2,Tanas3,gerry}. In \cite{mesina1,mesina2} it was proposed a scheme for producing a superposition of two arbitrary Glauber coherent states via parity and displacement operator. The other scheme for preparation of superposition of coherent states in cavity QED was studied in \cite{Zeng,zhen} in which an atom either flying through or trapped within a cavity, is controlled by the classical Stark effect. ECS's have many applications in quantum optics and quantum information processing \cite{Cochrane,Oliveira,Kim,Milburn,Munro,wang3,wang5,wang1,Vogel1,Salimi}. In \cite{wang2,wang4} the required conditions for the maximal entangled states of the form $|\psi\rangle=\mu|\alpha\rangle|\beta\rangle+\nu|\gamma\rangle|\delta\rangle$ have been studied and then have been generalized to the state $ |\psi\rangle=\mu|\alpha\rangle|\beta\rangle+\lambda|\alpha\rangle|\delta\rangle+
 \rho|\gamma\rangle|\beta\rangle+\nu|\gamma\rangle|\delta\rangle$  in Ref. \cite{najarbashi}. Generation of multipartite ECS's and entanglement of multipartite states constructed by linearly independent coherent states are investigated in \cite{Barry,Enk1}. In \cite{najarbashi1} it was considered the production and entanglement properties of the generalized balanced N-mode Glauber coherent states of the form
\be\label{mainstate}
|\Psi^{(d)}_{N}\rangle=\frac{1}{\sqrt{M^{(d)}_N}}\sum_{i=0}^{d-1}\mu_i|\underbrace{\alpha_i\rangle\cdots|\alpha_i\rangle}_{\mathrm{N \ modes}},
\ee
which is a general form of  the balanced two-mode entangled coherent state $|\Psi^{(2)}\rangle_{bal}=\frac{1}{\sqrt{M^{(2)}}}(|\alpha\rangle|\alpha\rangle+\mu|\beta\rangle|\beta\rangle)$ and $|\Psi^{(3)}\rangle_{bal}=\frac{1}{\sqrt{M^{(3)}}}(|\alpha\rangle|\alpha\rangle+\mu_1|\beta\rangle|\beta\rangle+\mu_2|\gamma\rangle|\gamma\rangle)$. By assumption that the coherent states are linearly independent, these states recast in two qubit and qutrit form respectively. Then the entanglement of this states was evaluated by concurrence measure.
\par
Another problem which has been investigated extensively in quantum information processing is noise effect or decoherence which arise from the coupling of the system to its surroundings \cite{Enk1,Yao}. van Enk in \cite{Enk} introduced
the effect of noise on coherent states with the modes $1$ or $2$ after traveling through a noisy channel as
$|\alpha\rangle_{1(2)}|0\rangle_E\rightarrow|\sqrt{\eta}\alpha\rangle_{1(2)}|\sqrt{1-\eta}\alpha\rangle_E$
where the second state now refers to the environment and $\eta$ is the noise parameter, which
gives the fraction of photons that survives the noisy channel. The effect of noise  on entanglement between modes $1$ and $2$ in qubit and qutrit like ECS's was investigated in Ref.\cite{najarbashi2}.
\par
In this paper we use the displacement operator together with parity operation to construct the superposition of two coherent states $|\alpha\rangle+\mu|\beta\rangle$. By transmitting this superposition from 50-50 beam splitter the two-mode qubit like ECS, $|\Psi^{(2)}\rangle$, is generated. Moreover we introduce a method for producing qutrit like ECS using atom-field interaction in cavity QED and beam splitter. The Wigner functions for reduced density matrices of both two-mode qubit and qutrit like ECS's yield some information about the quantum entanglement of ECS's. This information comes from the separation of peaks of Wigner quasi-probability  as a function
of ECS's parameters (e.g., $\alpha , \beta , \mu$ in state $|\alpha\rangle+\mu|\beta\rangle$). The results are compared with another useful measure such as concurrence which confirm the outcomes.
Moreover we investigate the noise effects on Wigner function. As a result   Wigner function is affected by noise effect and the entanglement of ECS's is decreased after traveling through the noisy channel.
\par
The outline of this paper is as follows: In section 2 we propose a method for generation of two-mode ECS's and investigate one-mode Wigner quasi-probability distribution function. We compare the Wigner function with bipartite entanglement which is obtained by concurrence measure. Section 3 devoted to the qutrit like ECS generated via cavity QED and beam splitter. The effect of noise on one-mode Wigner function of ECS is studied in section 4. Our conclusions are summarized in section 5.
\section{Generation of Two-mode Qubit like ECS}
In this section, we consider how to generate the two-mode ECS. The first problem is the generation of the superposition of two number coherent states like as $|\alpha\rangle+\mu|\beta\rangle$, (up to normalization factors). Various attempts proposed scheme for generating of discrete superposition of coherent states \cite{Sanders3,Milburn1,Milburn2,Enk1,Cheong1, Yurke1,Yurke2}.
\subsection{Generation of Two-mode Qubit like ECS}
\begin{figure}[ht]
\centerline{\includegraphics[width=8cm]{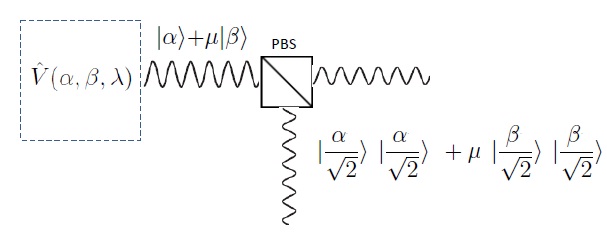}}
\caption{\small {Experimental set up for generating ECS.} \label{setup} }
\end{figure}

 In \cite{mesina1,mesina2}, the authors use the displacement operator together with parity operation to construct the unitary operation
\be
\hat{U}(\lambda,\alpha)=e^{i\lambda{\hat{D}}(\alpha)\hat{\Pi}},
\ee
where $\lambda$ is real number, $\hat{\Pi}=\cos(\pi \hat{a}^{\dag} \hat{a})$ is a Hermitian and unitary operator with property $\hat{\Pi}|n\rangle=(-1)^{n}|n\rangle$ where $|n\rangle$ refers to photon number state and $\hat{D}(\alpha)=exp(\alpha\hat{a}^{\dag}-\alpha^*\hat{a})$ is usual displacement operator. In order to obtain a linear combination of two arbitrary Glauber coherent states, it is enough  to use another displacement operator, $\hat{D}(\beta)$:
\be
\hat{D}(\beta)\hat{U}(\lambda,\alpha)|0\rangle=\cos\lambda|\beta\rangle+i\sin\lambda|\alpha+\beta\rangle.
\ee
Using $\hat{V}(\alpha,\beta,\lambda)=\hat{D}(\alpha)\hat{U}(\lambda,\beta-\alpha)$, one may recast the above state in a convenient form as follows
\be\label{1}
\hat{V}(\alpha,\beta,\lambda)|0\rangle=\cos\lambda|\alpha\rangle+i\sin\lambda e^{iIm(\alpha\beta^{*})}|\beta\rangle.
\ee
We  next  use  polarizing beam splitter (PBS). The polarizing beam splitter is commonly made by cementing together two birefringent materials like calcite or quartz, and has the property of splitting a light beam into its orthogonal linear polarizations.
The beam splitter interaction given by the unitary transformation
\be
\hat{B}_{i-1,i}(\theta)=\exp[\theta(\hat{a}_{i-1}^{\dag}\hat{a}_{i}-\hat{a}_{i}^{\dag}\hat{a}_{i-1})],
\ee
which $\hat{a}_{i-1}$, $\hat{a}_{i}$, $\hat{a}^{\dag}_{i-1}$ and $\hat{a}^{\dag}_{i}$ are the annihilation and creation operators of the field mode $i-1$ and $i$, respectively. Using Baker-Hausdorf formula, the action of the beam splitter on two modes $i-1$ and $i$ ,  can be expressed as
\be\label{action}
\hat{B}_{i-1,i}(\theta)\left(
  \begin{array}{c}
    \hat{a}_{i-1} \\
    \hat{a}_{i} \\
  \end{array}
\right)\hat{B}^{\dag}_{i-1,i}(\theta)=\left(
  \begin{array}{c}
    \hat{a}'_{i-1} \\
    \hat{a}'_{i} \\
  \end{array}
\right)=\left(
          \begin{array}{cc}
            \cos\theta & -\sin\theta \\
            \sin\theta & \cos\theta \\
          \end{array}
        \right)\left(
                 \begin{array}{c}
                   \hat{a}_{i-1} \\
                   \hat{a}_{i} \\
                 \end{array}
               \right).
\ee
For $50-50$ beam splitter, i.e. $\theta=\pi/4$ the above equation reduces to
 \be
\hat{B}_{1,2}(\pi/4)|\alpha'\rangle_{_{1}}|0\rangle_{_{2}}=|\frac{\alpha'}{\sqrt{2}}\rangle_{_{1}}|\frac{\alpha'}{\sqrt{2}}\rangle_{_{2}}.
\ee
This  result says that like classical light wave where the incident intensity is evenly divided between the two output beams, e.g. half the incident average photon number, $\frac{|\alpha|^{2}}{2}$, emerges in each beam. Note that the output is not entangled. For producing ECS suppose that our input state is a superposition of two coherent states as $|\alpha'\rangle_{_{1}}+\mu|\beta'\rangle_{_{1}}$.  Following the procedure above, we may then, obtain the output state (see figure \ref{setup})
\be
\hat{B}_{1,2}(\pi/4)(|\alpha'\rangle_{_{1}}+\mu|\beta'\rangle_{_{1}})\otimes|0\rangle_{_{2}}=|\frac{\alpha'}{\sqrt{2}}\rangle_{_{1}}|\frac{\alpha'}{\sqrt{2}}\rangle_{_{2}}+
\mu|\frac{\beta'}{\sqrt{2}}\rangle_{_{1}}|\frac{\beta'}{\sqrt{2}}\rangle_{_{2}}.
\ee
By renaming $\alpha\equiv\frac{\alpha'}{\sqrt{2}}$ and $\beta\equiv\frac{\beta'}{\sqrt{2}}$ we have
$
|\alpha\rangle_{_{1}}|\alpha\rangle_{_{2}}+\mu|\beta\rangle_{_{1}}|\beta\rangle_{_{2}},
$
which is a two-mode qubit like ECS.
The term qubit like state comes from the fact that two coherent states $|\alpha\rangle$ and $|\beta\rangle$ are in general nonorthogonal and linearly independent, hence they may span a two dimensional  Hilbert space i.e. $\{ |0\rangle, |1\rangle\}$, defined as
\be
\begin{array}{l}
\left| \alpha  \right\rangle  = \left| 0 \right\rangle ,\,\\
\left| \beta  \right\rangle  = {N_1}\left| 1 \right\rangle  + {p_1}\left| 0 \right\rangle ,
\end{array}
\ee
in which ${N_1} = \sqrt {1 - {p_1}^2}$ and $p_1=\langle\alpha|\beta\rangle$. Therefore the state $\left| {{\Psi ^{(2)}}}\right\rangle$ can be recast in   two qubit form
\be
\left| {{\Psi ^{(2)}}} \right\rangle  = \frac{1}{{\sqrt {{M^{(2)}}} }}\left\{ {(1 + \mu {p_1}^2)\left| {00} \right\rangle  + \mu {N_1}{p_1}(\left| {10} \right\rangle  + \left| {01} \right\rangle ) + \mu {N_1}^2\left| {11} \right\rangle } \right\},
\ee
where $M^{(2)}=1+\mu^2+2\mu Re(p^2)$ is normalization factor.
So the state $\left| {{\Psi ^{(2)}}} \right\rangle$  is a state with two constituent modes, where each of them are defined in two dimensional Hilbert space. Note that we used the superscript (2) for qubit-like states to distinguish it from that of qutrit like states which will be defined in the next section.
\subsection{One-Mode Wigner Function for Qubit like ECS's}
\begin{figure}[ht]
\centerline{\includegraphics[width=12cm]{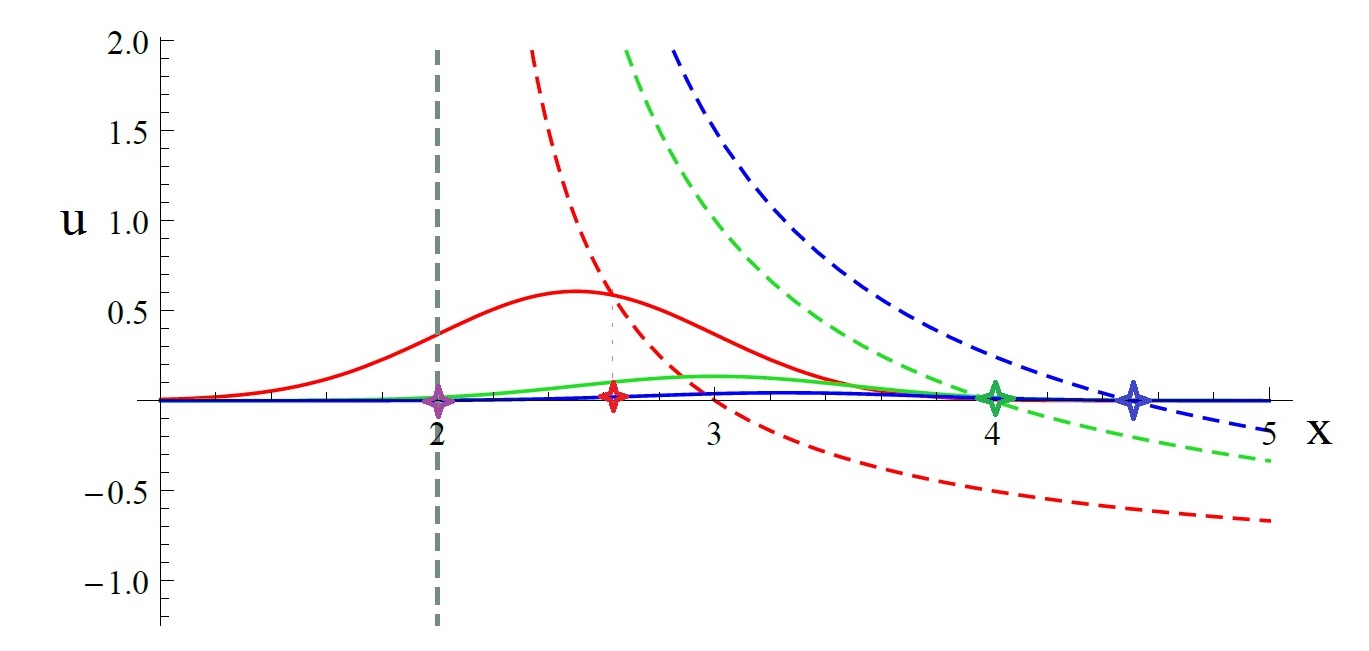}}
\caption{\small {(Color online) $u=e^{-(x-\alpha)^2+(x-\beta)^2}$ (full line) and $u=\frac{x-\beta}{\alpha-x}$ (dashed line) as function of $x$ for given $\alpha=2$: (a)$\beta=3$ (Red), (b)$\beta=4$ (Green) and (c)$\beta=4.5$ (Blue).} \label{s} }
\end{figure}

\begin{figure}[ht]
\centerline{\includegraphics[width=10cm]{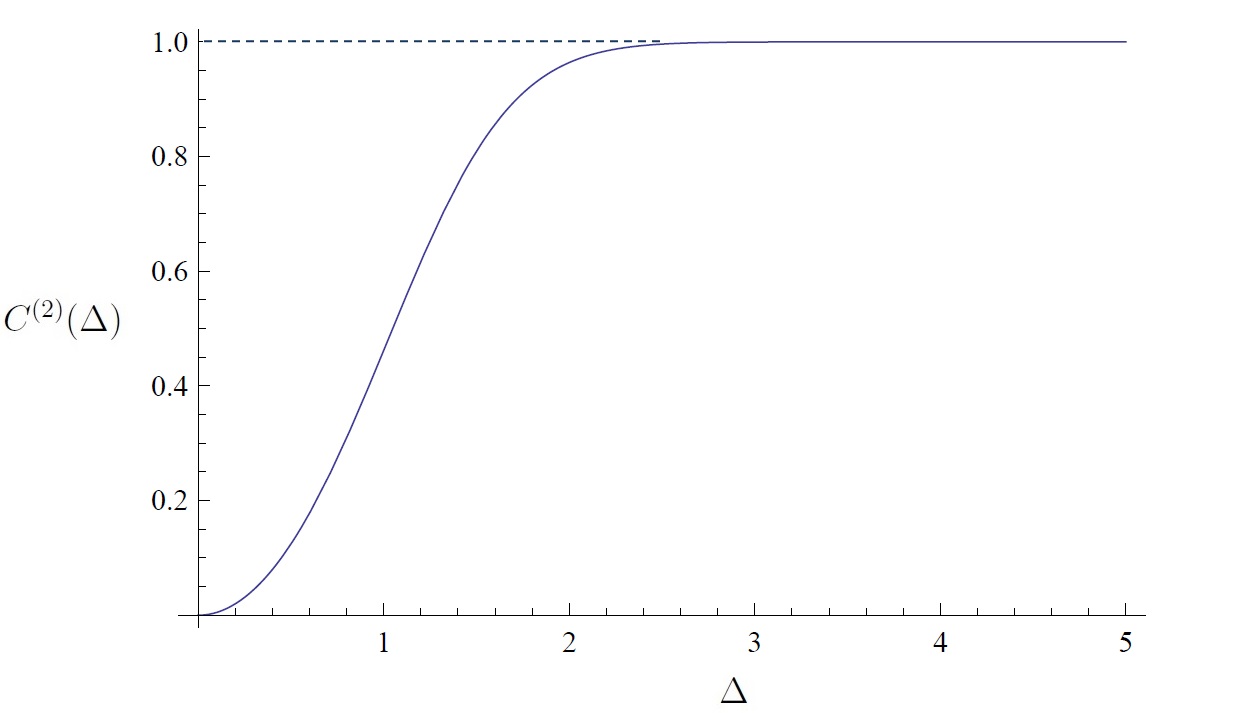}}
\caption{\small {(Color online) Concurrence of $|\Psi^{(2)}\rangle$ as a function of separation of two peaks $\Delta$.} \label{sep} }
\end{figure}

One of the important quasi-probability distribution over phase
space is the Wigner function. The Wigner function seems to be the earliest
introduced of the phase-space quasi-probability distributions, making its debut
in 1932. Wigner function is defined as \cite{Book}
\be\label{wigner}
W(\gamma)=\frac{1}{\pi^2}\int d^2\lambda C_W(\lambda)e^{\lambda^*\gamma-\lambda\gamma^*},
\ee
in which $C_W(\lambda)=Tr(\rho D(\lambda))$ is one mode Wigner characteristic function and $D(\lambda)$ is displacement operator. $\rho$ is reduced density matrix which is obtained by partially tracing out second mode. Here we investigate the Wigner function behaviour for reduced density matrix of qubit like ECS.\\
Let us consider qubit like ECS which was generated in previous section in the form
\be\label{si}
|\Psi^{(2)}\rangle=\frac{1}{\sqrt{M^{(2)}}}(|\alpha\rangle|\alpha\rangle+\mu|\beta\rangle|\beta\rangle),
 \ee
in which for simplicity we assume that $\alpha,\beta$ and $\mu$ are real parameters and $\mu=1$. The reduced density matrix is
\be
\rho=\frac{1}{M^{(2)}}\{|\alpha\rangle\langle\alpha|+\mu p(|\alpha\rangle\langle\beta|+|\beta\rangle\langle\alpha|)+\mu^2|\beta\rangle\langle\beta|\}.
\ee
Hence Wigner characteristic function reads
\be
C^{(2)}_W(\lambda)=\frac{1}{M^{(2)}}\{\langle\alpha|D(\lambda)|\alpha\rangle+\mu^2\langle\beta|D(\lambda)|\beta\rangle+\mu p(\langle\beta|D(\lambda)|\alpha\rangle+\langle\alpha|D(\lambda)|\beta\rangle)\},
\ee
by substituting $C^{(2)}_W(\lambda)$ in Eq.(\ref{wigner}), the corresponding Wigner function is obtained as
\be
W^{(2)}(\gamma)=\frac{2}{\pi M^{(2)}}\{e^{-2|\gamma-\alpha|^2}+\mu^2e^{-2|\gamma-\beta|^2}+\mu e^{-(|\alpha|^2+|\beta|^2)}e^{-2|\gamma|^2}
(e^{2(\gamma\beta^*+\gamma^*\alpha)}+e^{2(\gamma\alpha^*+\gamma^*\beta)})\}.
\ee
Setting $\gamma=x+iy$, we plot diagram of $W^{(2)}(x,y)$ as a function of $x$ and $y$ for given $\alpha$ and $\beta$.
\begin{figure}[ht]
\centerline{\includegraphics[width=14cm]{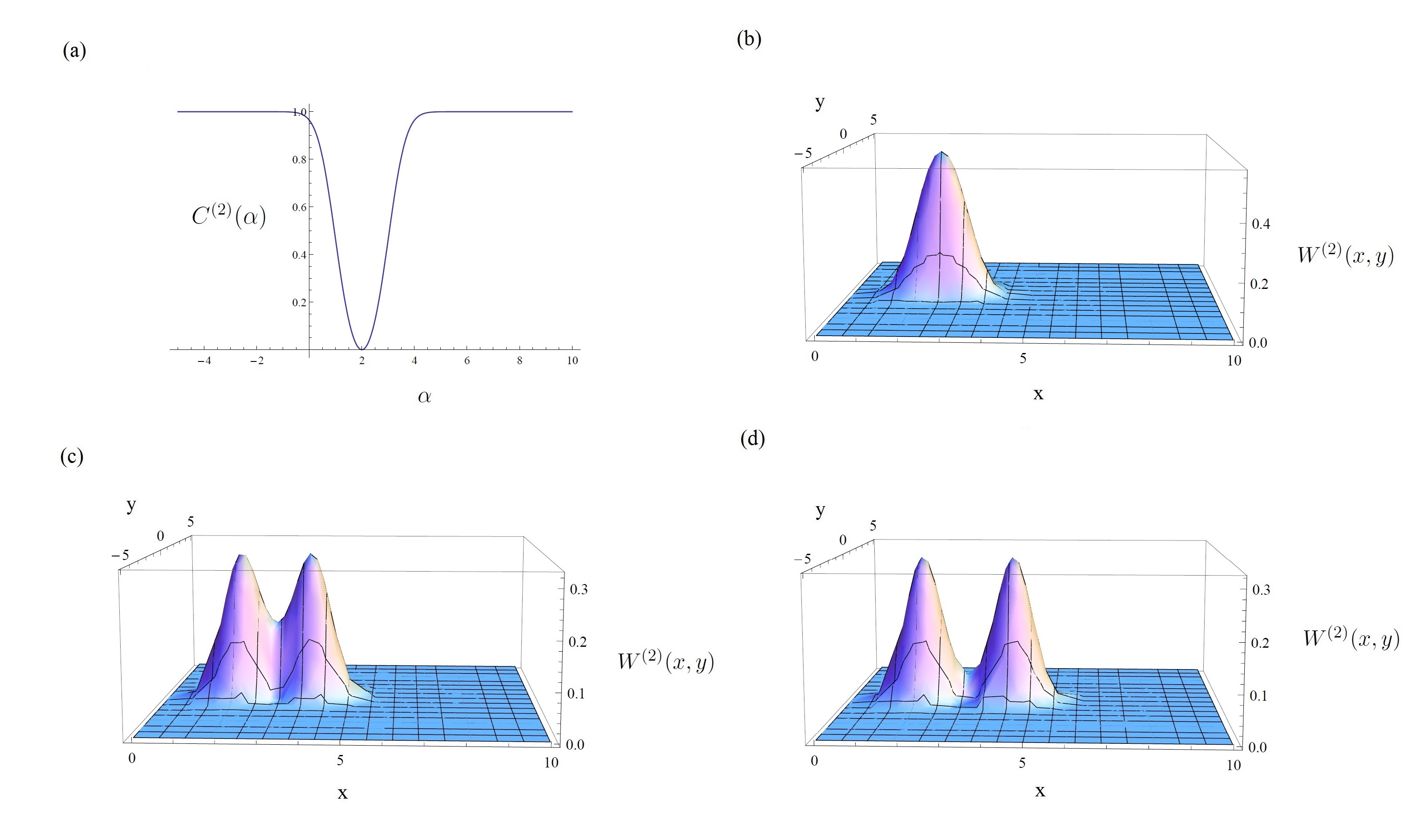}}
\caption{\small {(Color online) (a) Concurrence of $|\Psi^{(2)}\rangle$ as a function of $\alpha$ for a given $\beta=2$ and $W^{(2)}(x,y)$ as function of $x$ and $y$ for $\alpha=2$: (b) $\beta=2$, (c) $\beta=4$ and (d) $\beta=4.5$.} \label{wi} }
\end{figure}
Figure \ref{wi} shows that by considering $\alpha$ and $\beta$ as real numbers, the Wigner function moves along $x$ axis (real part of $\gamma$) while if we regard that $\alpha$ and $\beta$ are imaginary numbers, the Wigner function moves along $y$ axis (imaginary part of $\gamma$). Moreover, the study of Wigner function for reduced density matrix reveals information on the entanglement between modes in two-mode ECS, Eq.(\ref{si}). For more details, one can do a comparison between$(a),(b),(c)$ and $(d)$   in figure \ref{wi} which shows that if $\alpha=\beta$ the state is separable and there is one peak (figure (\ref{wi}b)). For $\alpha\neq\beta$ the state is entangled. As a result, the separation of two peaks can be a witness for amount of entanglement of $|\Psi^{(2)}\rangle$. For clarity we assume that $y=0$ and solve the equation $\frac{dW^{(2)}(x)}{dx}=0$:
\be\label{2}
(\alpha-x)e^{-2(\alpha-x)^2}+(\beta-x)e^{-2(\beta-x)^2}+(\alpha+\beta-2x)e^{-(\alpha-x)^2-(\beta-x)^2}=0.
\ee
Clearly, $x=\alpha=\beta$ is one solution for Eq.(\ref{2}), which means that the diagram of $W^{(2)}(\gamma)$ assumes  one peak at point $x=\alpha$. Whereas if $\alpha\neq\beta$ and considering $u=e^{-(x-\alpha)^2+(x-\beta)^2}$, Eq.(\ref{2}) is rewritten as
\be
(\alpha-x)u^2+(-2x+\alpha+\beta)u+\beta-x=0,
\ee
which has two solutions $u=-1$ and $u=\frac{x-\beta}{\alpha-x}$. The former must be ignored as $u$ is always positive.
The equation $u=e^{-(x-\alpha)^2+(x-\beta)^2}=\frac{x-\beta}{\alpha-x}$ is a transcendent equation and it can not be solved  in an algebraic way. However, if we plot  both sides of this equation for a given $\alpha$ and $\beta$, then their intersections are required solutions (see  figure \ref{s}).
Figure \ref{s} shows that by  increasing $\Delta=\alpha-\beta$ the separation of two maximum in Wigner function  increases too. This comes from the fact that  putting  $x=\alpha$ in   Eq. (\ref{2}) yields $(\beta-\alpha)[e^{-2(\beta-\alpha)^2}+e^{-(\beta-\alpha)^2}]\simeq0$ which is satisfied only for $|\alpha-\beta|\gg 1$.
The result is that the separation of two peaks in Wigner function is a monotone function of $\Delta$ and  for large $\alpha-\beta$, the separation of two peaks in Wigner function  just depends on  $\Delta=\alpha-\beta$.
To confirm the above  results we use another measure, the so called concurrence, which illustrates that the entanglement of $|\Psi^{(2)}\rangle$, depends solely on $\Delta$.
As in general  two coherent states $|\alpha\rangle$ and $|\beta\rangle$ are linearly independent,  they may span a two dimensional qubit like Hilbert space $\{|0\rangle,|1\rangle\}$, hence the two-mode coherent state $|\Psi^{(2)}\rangle$ can be recast in two qubit form \cite{najarbashi1}.
 For any pure state in the form $|\psi\rangle=a_{00}|00\rangle+a_{01}|01\rangle+a_{10}|10\rangle+a_{11}|11\rangle$, the concurrence is defined as $C=2|a_{00}a_{11}-a_{01}a_{10}|$ \cite{Wootters1,Wootters2}. Therefore the concurrence of Eq.(\ref{si}) in term of $\Delta$ is rewritten as
\be\label{3}
C^{(2)}(\Delta)=\frac{1-e^{-\Delta^2}}{1+e^{-\Delta^2}}.
\ee
This equation shows that the concurrence is a monotone function of $\Delta$. If the separation of two peaks, becomes large ($\Delta\rightarrow\infty$), the concurrence tends to its maximum ($C_{max}^{(2)}=1$), while for small separation (i.e. $\Delta\rightarrow0$) the concurrence tends to zero and the state becomes separable. Figure \ref{sep} shows  the concurrence as a monotone function of separation of two peaks (see figure \ref{sep}).\\
\section{Generation of Two-mode Qutrit like ECS}
In the previous section, we introduced a method to produce superposition of even number of coherent states. These states lead to qubit like ECS. One may construct qutrit like ECS. To this aim, we introduce a controllable method for producing qutrit like ECS. Then the Wigner function for reduced density matrix and bipartite entanglement of this state is investigated.
\subsection{Generation of Two-mode Qutrit like ECS}
 Here we use atom-field interaction in cavity QED and beam splitter to produce qutrit like ECS. The hamiltonian of Jaynes-Cummings model \cite{Zeng,zhen} is defined as
\be
\hat{H}_{int}=\omega_{c}\hat{a}^\dag \hat{a}+\omega_{a}\hat{\sigma}_{z}+g(\hat{a}^\dag \hat{\sigma}^{-}+\hat{a}\hat{\sigma}^{+})+\varepsilon e^{-i\omega_{L}t}\hat{\sigma}^{+}+\varepsilon^{*} e^{i\omega_{L}t}\hat{\sigma}^{-},
\ee
where $\hat{a}(\hat{a}^{\dag})$ is the annihilation (creation) operator for the cavity field and for simplicity we assume that $\hbar=1$. $\omega_c$, $\omega_L$ and $\omega_a$ are the frequencies of the cavity, classical field and the atomic transition frequency between the excited state $|e\rangle$ and the ground state $|g\rangle$ respectively. $g$ is the atom-cavity coupling constant. The complex amplitude is represented by $\varepsilon$. $\hat{\sigma}^{\pm}$ and $\hat{\sigma}_{z}$ are the atomic transition operators which are given by
\be
\begin{array}{l}
\hat{\sigma}^{+}=|e\rangle\langle g|,
\hat{\sigma}^{-}=|g\rangle\langle e|,\\
\hat{\sigma}_{z}=|e\rangle\langle e|-|g\rangle\langle g|.
\end{array}
\ee
In this method, an atom interact alternately with a (resonant) classical field and with the (dispersive) cavity field. Let us suppose that the atom is not affected by the cavity field and is initially resonant with the classical field so the interaction hamiltonian is given by $\hat{H}_{I1}=\varepsilon e^{-i\varphi}\hat{\sigma}^{+}+\varepsilon^{*} e^{i\varphi}\hat{\sigma}^{-}$. If the atom is initially in the ground state $|g\rangle$ or $|e\rangle$, after the atom interacts with the classical field we have
\be
\begin{array}{l}
|g\rangle \rightarrow \frac{1}{\sqrt{1+|\varepsilon_{k}|^2}}(|g\rangle+\varepsilon_{k}|e\rangle),\\
|e\rangle\rightarrow \frac{1}{\sqrt{1+|\varepsilon_{k}|^2}}(-\varepsilon_{k}^{*}|g\rangle+|e\rangle),
\end{array}
\ee
respectively, where $\varepsilon_{k} (k=0,1,2,...)$ is an adjustable complex number controlled by the parameters of the classical field. Now if the atom is interacting dispersively with the cavity field and far away from the classical field, the effective hamiltonian of the atom-cavity system is given by $\hat{H}_{I2}=\frac{g^2}{\widetilde{\Delta}}\hat{a}^{\dag} \hat{a}\hat{\sigma}_z$, in which $\widetilde{\Delta}=\omega_a-\omega_c$. We assume that the cavity field is initially in a coherent state $|\alpha\rangle$, thus the total system has the form $|\psi_1\rangle\equiv \frac{1}{\sqrt{1+|\varepsilon_{0}|^2}}(|g\rangle|\alpha\rangle+\varepsilon_0|e\rangle|\alpha\rangle)$. After an interaction time $t =\frac{\pi\widetilde{\Delta}}{2g^2} $, the atom-cavity system evolves to
\be
|\psi_1\rangle\rightarrow \frac{1}{\sqrt{1+|\varepsilon_{0}|^2}}(|g\rangle|i\alpha\rangle+\varepsilon_{0}|e\rangle|-i\alpha\rangle).
\ee
Again the atomic transition is resonant with the classical field but far away from the cavity field. After a given time and performing a measurement on the atom we have
\be\label{7}
|\psi_1'\rangle=\frac{1}{\sqrt{M^{(2)}}}(-\varepsilon_0\varepsilon_1^*|-i\alpha\rangle+|i\alpha\rangle),
\ee
where $M^{(2)}=1+|\varepsilon_0\varepsilon_1^*|^2-2Re(\varepsilon_0\varepsilon_1^*)e^{-2|\alpha|^2}$. Further the phase shifter $\hat{\mathcal{P}}=e^{-i\frac{\pi}{2}\hat{a}^{\dag}\hat{a}}$ transforms the state Eq.(\ref{7}) as
\be\label{9}
\hat{\mathcal{P}}|\psi_1'\rangle=\frac{1}{\sqrt{M^{(2)}}}(-\varepsilon_0\varepsilon_1^*|\alpha\rangle+|-\alpha\rangle).
\ee
Similar to the first step $N=1$, once again the atom interacts alternately with a (resonant) classical field and with the (dispersive) cavity field $N=2$, then by performing a measurement on the atom in ground state, we can generate superpositions of three coherent states with adjustable weighting factors as the form
\be\label{3}
|\psi'_2\rangle=\frac{1}{\sqrt{M^{(3)}}}(|2\alpha\rangle-(\varepsilon_0\varepsilon_2^*+\varepsilon_0\varepsilon_1^*)|0\rangle-\varepsilon_1\varepsilon_2^*|-2\alpha\rangle),
\ee
in which $M^{(3)}$ is normalization factor. In \cite{Zeng} the optimal values of $\varepsilon_0$, $\varepsilon_1$, and $\varepsilon_2$ determined to be $-0.8200$, $2.1184$, and $-0.4720$, respectively. Acting displacement operator $\hat{D}(\beta)=e^{\beta \hat{a}^{\dag}-\beta^*\hat{a}}$ with property $\hat{D}(\beta)|\alpha\rangle=|\alpha+\beta\rangle$, onto state Eq.(\ref{3}) leads to the following superposition of coherent states
\be\label{8}
\hat{D}(\beta)|\psi'_2\rangle=\frac{1}{{\sqrt{2e^{-8\alpha^2}+5.4e^{-2\alpha^2}+3.8225}}}(|2\alpha+\beta\rangle+1.35|\beta\rangle+|-2\alpha+\beta\rangle).
\ee
By transmitting this state Eq.(\ref{8}) through $50-50$ beam splitter we have
\be\label{QED}
|\Phi^{(3)}\rangle=\frac{1}{\sqrt{M^{(3)}}}(|\frac{2\alpha+\beta}{\sqrt{2}}\rangle|\frac{2\alpha+\beta}{\sqrt{2}}\rangle
+1.35|\frac{\beta}{\sqrt{2}}\rangle|\frac{\beta}{\sqrt{2}}\rangle+|\frac{-2\alpha+\beta}{\sqrt{2}}\rangle|\frac{-2\alpha+\beta}{\sqrt{2}}\rangle).
\ee
In the following, we concern the one-mode Wigner function of qutrit like ECS's. Firstly, we will consider the general qutrit like ECS state $|\Psi^{(3)}\rangle=\frac{1}{\sqrt{M'^{(3)}}}(|\alpha\alpha\rangle+\mu_1|\beta\beta\rangle+\mu_2|\gamma\gamma\rangle)$ then as an especial example we will consider the state Eq.(\ref{QED}). One can show that the state $|\Psi^{(3)}\rangle$ is a two-mode qutrit like ECS. To this end we assume the set $\{ |\alpha\rangle,|\beta\rangle, |\gamma\rangle \}$ are in general linearly independent, i.e. they span the three dimensional Hilbert space  $\{ |0\rangle, |1\rangle\,|2\rangle \}$.
Therefore we can define three orthonormal basis as
\be\label{newbasis}
\begin{array}{l}
|0\rangle=|\alpha\rangle,\\
|1\rangle=\frac{1}{\sqrt{1-p_{1}^{2}}}(|\beta\rangle-p_{1}|\alpha\rangle),\\
|2\rangle=\sqrt{\frac{1-p_{1}^{2}}{1-p^2_{1}-p^2_{2}-p^2_{3}+2p_1p_2p_3}}\left(|\gamma\rangle+(\frac{p_{1}p_{3}-p_{2}}{{1-p_{1}^{2}}})|\beta\rangle+(\frac{p_{1}p_{2}-p_{3}}{{1-p_{1}^{2}}})|\alpha\rangle\right),
\end{array}
\ee
where $p_{1}=\langle\alpha|\beta\rangle$,  $p_{2}=\langle\gamma|\beta\rangle$, $p_{3}=\langle\gamma|\alpha\rangle$
and again for simplicity we assumed that   all the parameters  are real.
By substituting Eq. (\ref{newbasis})  in state $|\Psi^{(3)}\rangle$ we have
\be
\begin{array}{l}
|\Psi^{(3)}\rangle=\frac{1}{\sqrt{M'^{(3)}}}\{(1+\mu_1p_1^2+\mu_2p_3^2)|00\rangle+N_1^2(\mu_1+\mu_2x^2)|11\rangle+\mu_2N_2^2|22\rangle\\
~~~~~~~~+N_1(\mu_1p_1-\mu_2xp_3)(|10\rangle+|01\rangle)-\mu_2xN_1N_2(|21\rangle+|12\rangle)\\
~~~~~~~~+\mu_2N_2p_3(|20\rangle+|02\rangle)\},
\end{array}
\ee
where $ x=\frac{p_{1}p_{3}-p_{2}}{{1-p_{1}^{2}}}$, $N_1=\sqrt{1-p_1^2}$, $N_2=\sqrt{1-p_3^2-x^2N_1^2}$ and
$M'^{(3)}=1+\mu_1^2+\mu_2^2+2\mu_1p_1^2+2\mu_1\mu_2p_2^2+2\mu_2p_3^2$ is normalization factor. This state is clearly a two qutrit like state.
\subsection{One-Mode Wigner Function for Qutrit like ECS's}
\begin{figure}[ht]
\centerline{\includegraphics[width=14cm]{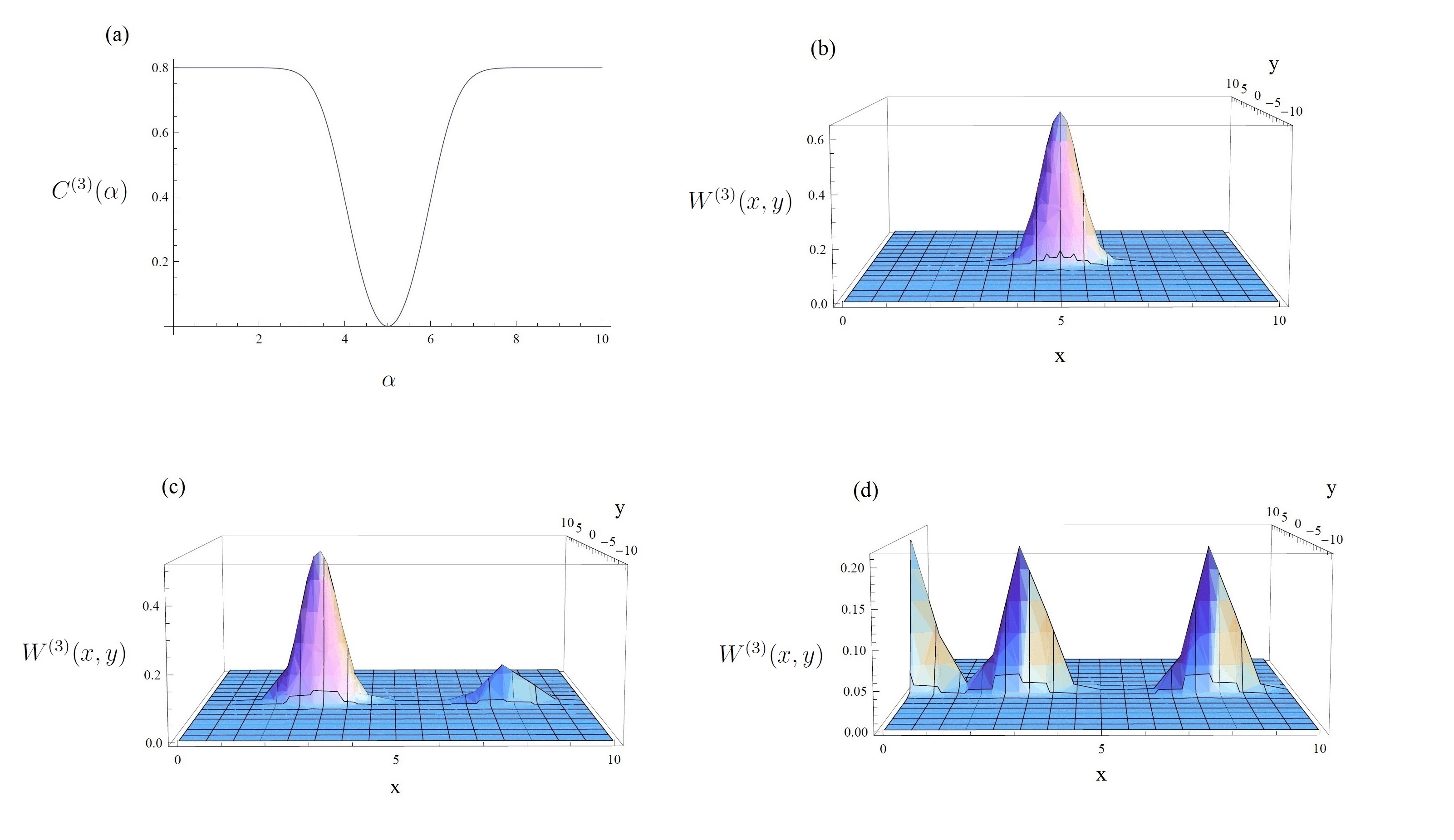}}
\caption{\small {(Color online) (a) Concurrence of $|\Psi^{(3)}\rangle$ as a function of $\alpha$ for $\beta=\gamma=5$
and $W^{(3)}(x,y)$ as function of $x$ and $y$: (b) $\alpha=\beta=\gamma=5$, (c) $\alpha=\beta=3, \gamma=8$, (d) $\alpha=0, \beta=3, \gamma=8$.} \label{cw} }
\end{figure}

\begin{figure}[ht]
\centerline{\includegraphics[width=13cm]{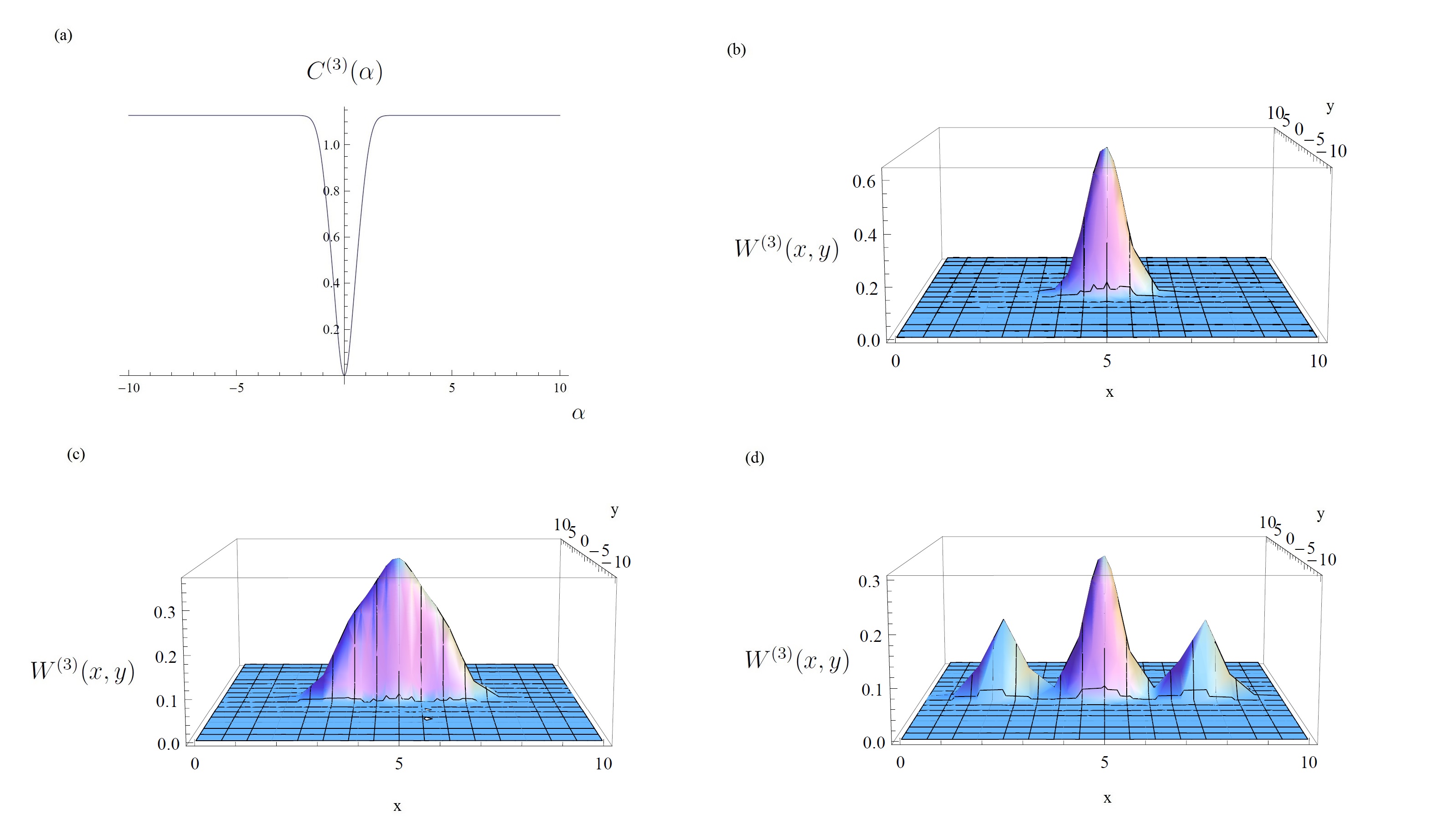}}
\caption{\small { (Color online) (a) Concurrence of $|\Phi^{(3)}\rangle$ as a function of $\alpha$ and $W^{(3)}(x,y)$ as function of $x$ and $y$ for $\beta=7$: (b) $\alpha=0$, (c) $\alpha=1$ and (d) $\alpha=2$.}} \label{qed}
\end{figure}

\begin{figure}[ht]
\centerline{\includegraphics[width=11cm]{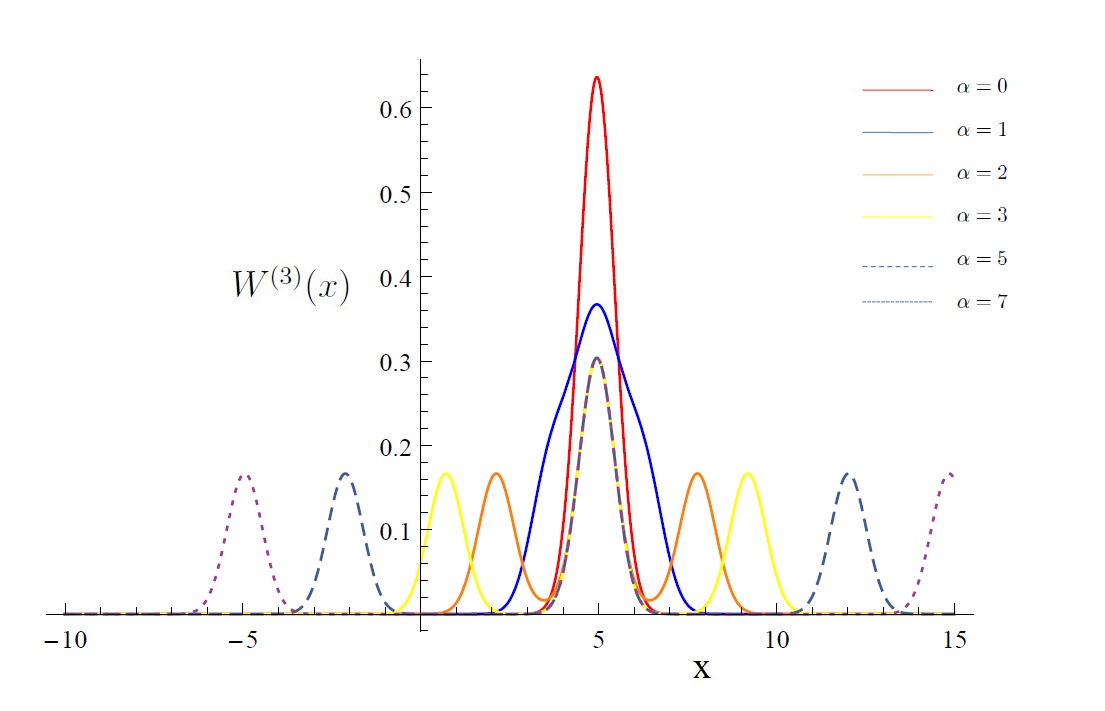}}
\caption{\small {(Color online) Wigner function of reduced density matrix of $|\Phi^{(3)}\rangle$ as a function of $x$ for $\beta=7$.}} \label{q}
\end{figure}

Let us consider two qutrit like ECS
\be\label{qutrit}
|\Psi^{(3)}\rangle=\frac{1}{\sqrt{M'^{(3)}}}(|\alpha\rangle|\alpha\rangle+\mu_1|\beta\rangle|\beta\rangle+\mu_2|\gamma\rangle|\gamma\rangle),
\ee
Again for simplicity we assumed that $\alpha,\beta$, $\gamma$, $\mu_1$ and $\mu_2$ are real parameters. Using the definition Eq.(\ref{wigner}) the Wigner function is obtained as
\be
\begin{array}{l}
W^{(3)}(\delta)=\frac{2}{\pi M'^{(3)}}\{e^{-2|\delta-\alpha|^2}+\mu_1^2e^{-2|\delta-\beta|^2}+\mu_2^2e^{-2|\delta-\gamma|^2}
+\mu_1p_1e^{-\frac{1}{2}(\alpha+\beta)^2}e^{-2|\delta|^2}(e^{2(\delta\beta+\delta^*\alpha)}+e^{2(\delta^*\beta+\delta\alpha)})\\
+\mu_1\mu_2p_2e^{-\frac{1}{2}(\gamma+\beta)^2}e^{-2|\delta|^2}(e^{2(\delta\gamma+\delta^*\beta)}+e^{2(\delta^*\gamma+\delta\beta)})
+\mu_2p_3e^{-\frac{1}{2}(\alpha+\gamma)^2}e^{-2|\delta|^2}(e^{2(\delta\gamma+\delta^*\alpha)}+e^{2(\delta^*\gamma+\delta\alpha)})\}.
\end{array}
\ee
Taking $\delta=x+iy$ and assuming $\mu_1=\mu_2=1$, the Wigner function $W^{(3)}(x,y)$ can be represented diagrammatically as a function of $x$ and $y$ for given $\alpha$, $\beta$ and $\gamma$ in the figure \ref{cw}.
One can  see  that for $\alpha=\beta=\gamma=5$ there is just one peak in diagram of $W^{(3)}(x,y)$ and the qutrit like ECS is reduced to  separable state ($C^{(3)}(\alpha)=0$), while for $\alpha\neq\beta\neq\gamma$ there exist three peaks and the state is in general entangled. Figure \ref{cw} shows that the distance of peaks in Wigner function $W^{(3)}(x,y)$ depends on parameters  $\Delta_{i}$ defined as
\be
\begin{array}{l}
\Delta_1=|\alpha-\beta|,\\
\Delta_2=|\alpha-\gamma|,\\
\Delta_3=|\beta-\gamma|,
\end{array}
\ee
which in turn implies that $W^{(3)}(x,y)$ is entanglement sensitive. To confirm this result we use the general concurrence measure
for bipartite state $|\psi\rangle=\sum_{i=1}^{d_{1}}\sum_{j=1}^{d_{2}}a_{ij}|e_{i}\otimes e_{j}\rangle$  \cite{Akhtarshenas}.
The norm of concurrence vector is obtained  as $C = 2 (\sum\limits_{i < j}^{d_1 } {\sum\limits_{k < l}^{d_2 } {\left| {a_{ik}a_{jl}-a_{il}a_{jk} } \right|^2 }})^{1/2}$, where $d_{1}$ and $d_{2}$ are dimensions of first and second part respectively. If the set $\{|\alpha\rangle,|\beta\rangle,|\gamma\rangle\}$ are linearly independent meaning they may span a three dimensional Hilbert space $\{|0\rangle,|1\rangle,|2\rangle\}$, hence  two mode coherent state $|\Psi^{(3)}\rangle$ can be recast in two qutrit form. Therefore the concurrence in terms of the separation of peaks is obtained as
\be
\begin{array}{l}
C^{(3)}(\Delta_1,\Delta_2,\Delta_3)=\frac{2}{3+2e^{-\Delta_1^2}+2e^{-\Delta_2^2}+2e^{-\Delta_3^2}}\{3+e^{-2\Delta_1^2}+e^{-2\Delta_2^2}+e^{-2\Delta_3^2}\\
~~~~~~~~~~~~~~~~~~~~~~~~+2e^{-\Delta_2^2-\Delta_3^2}-12e^{-\frac{1}{2}(\Delta_1^2+\Delta_2^2+\Delta_3^2)}+2e^{-\Delta_1^2}(e^{-\Delta_2^2}+e^{-\Delta_3^2})\}^{1/2}.
\end{array}
\ee
This equation shows that if all $\Delta_i\rightarrow0(i=1,2,3)$ the concurrence tends to zero and the state become non entangled and we have one peak (figure(\ref{cw}b)) while for $\Delta_i\rightarrow\infty$, the concurrence tends to its maximum value i.e. $C_{max}=1.154$ and three peaks appear as in figure (\ref{cw}d). We note that the Wigner function is positive for all values of $x$ and $y$.
\begin{figure}[ht]
\centerline{\includegraphics[width=11cm]{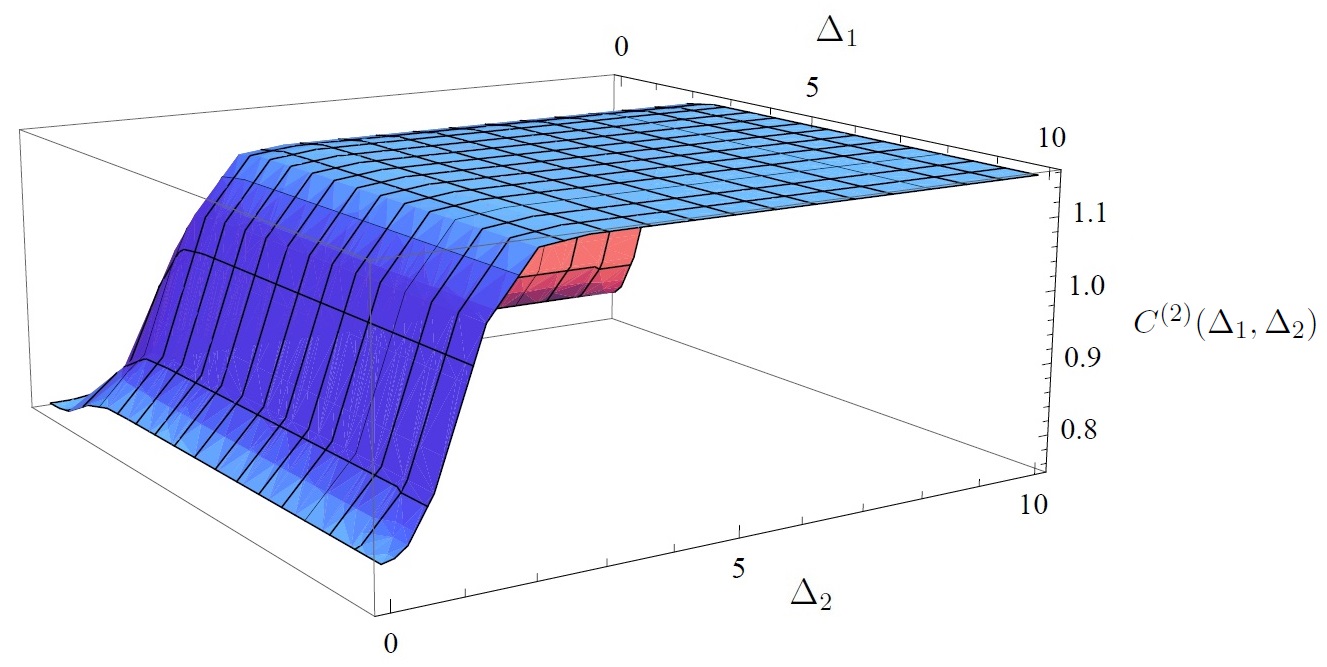}}
\caption{\small {(Color online) Concurrence of $|\Psi^{(3)}\rangle$ as function of $\Delta_1$ and $\Delta_2$ in the range $[0,10]$ for a given $\Delta_3=10$.}  }
\end{figure}
Another example for two-mode qutrit like ECS which is generated via atom-field interaction was given in Eq.(\ref{QED}). For this state, Wigner function for reduced density matrix is obtained as
\be
\begin{array}{l}
W^{(3)}(\delta)=\frac{2}{\pi M^{(3)}}\{e^{-2|\delta-\frac{2\alpha+\beta}{\sqrt{2}}|^2}+1.8225e^{-2|\delta-\frac{\beta}{\sqrt{2}}|^2}+e^{-2|\delta-\frac{-2\alpha+\beta}{\sqrt{2}}|^2}\\
~~~~~~~~~~~+1.35e^{-2\alpha^2-2\alpha\beta-\beta^2-2|\delta|^2}(e^{\sqrt{2}(\delta\beta+\delta^*(2\alpha+\beta)})+e^{\sqrt{2}(\delta^*\beta+\delta(2\alpha+\beta)})\\
~~~~~~~~~~~+1.35e^{-2\alpha^2+2\alpha\beta-\beta^2-2|\delta|^2}(e^{\sqrt{2}(\delta(-2\alpha+\beta)+\delta^*\beta})+e^{\sqrt{2}(\delta^*(-2\alpha+\beta)+\delta\beta})\\
~~~~~~~~~~~+e^{-4\alpha^2-\beta^2-2|\delta|^2}(e^{\sqrt{2}(\delta(-2\alpha+\beta)+\delta^*(2\alpha+\beta)})+e^{\sqrt{2}(\delta^*(-2\alpha+\beta)+\delta(2\alpha+\beta)})
\}.
\end{array}
\ee
Diagram of concurrence and Wigner function of reduced density matrix of the state $|\Phi^{(3)}\rangle$ is shown in figure \ref{qed}.
Figure (\ref{qed}a) shows that only for $\alpha=0$ the concurrence is zero and  independent of  $\beta$. This corresponds exactly with one peak in $W^{(3)}(x,y)$. By increasing $\alpha$ the width of the peak is increased so for $\alpha=2$, we see three peaks in diagram (see figure (\ref{qed}c) and (\ref{qed}d)). Moreover by assuming $y=0$ for given $\alpha$ and $\beta$, we draw a profile of $W^{(3)}(x)$ as a function of $x$ for given $\alpha$ and $\beta$ (see also figure \ref{q}).
Figure \ref{q} shows that if $\alpha=0$ Wigner function has one peak and the state is separable and by increasing $\alpha$, the peaks go away from each other and concurrence is raised.
\section{Noise Effect on One-Mode Wigner Function of ECS's}
\begin{figure}[ht]
\centerline{\includegraphics[width=15cm]{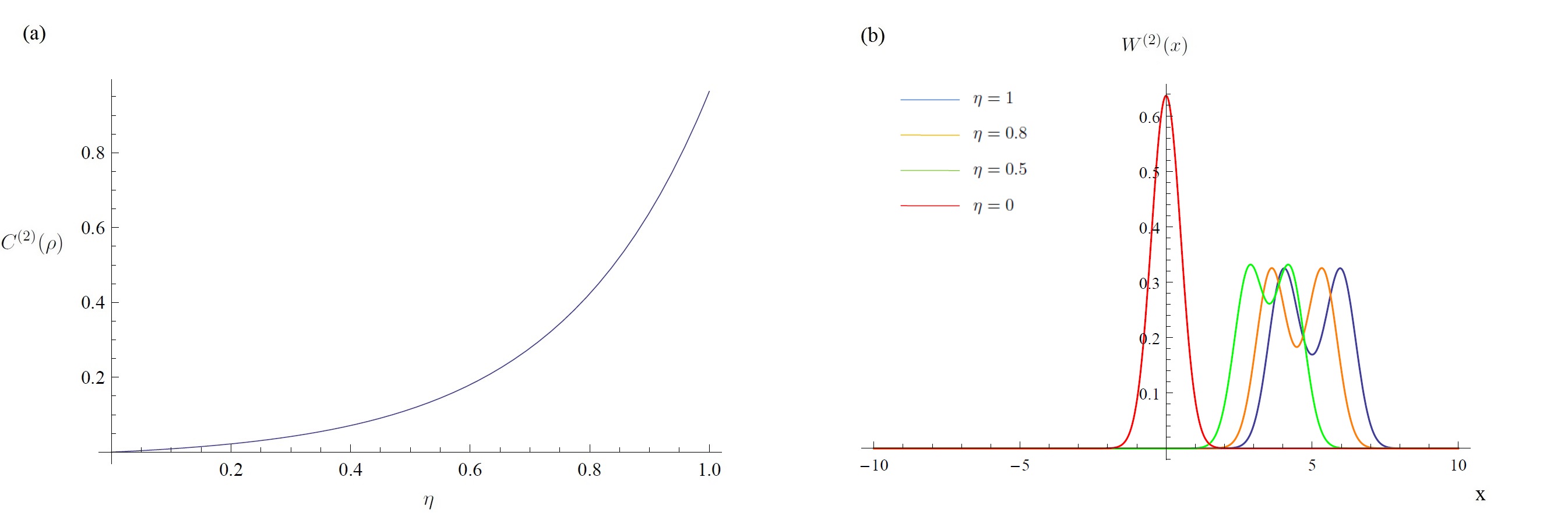}}
\caption{\small {(Color online) (a) $C^{(2)}(\rho)$ as a function of $\eta$, (b) $W^{(2)}(x)$ as a function of $x$ for a given $\eta$ and $\alpha=4,\beta=6$.} \label{NW} }
\end{figure}

Let us assume that the mode $1$ travel through a noisy channel characterized by
\be
|\alpha\rangle_{1}|0\rangle_E\rightarrow |\sqrt{\eta}\alpha\rangle_{1}|\sqrt{1-\eta}\alpha\rangle_E,
\ee
where the second state now refers to the environment and $\eta$ is the noise parameter, which gives the fraction of photons that survives the noisy channel \cite{Enk,Yao}. Here the effect of noise on Wigner function for reduced density matrix of qubit like ECS is investigated.
Let us consider the state $|\Psi^{(2)}\rangle$. After traveling through the noisy channel the state $|\Psi^{(2)}\rangle$ becomes
\be
|\Psi'^{(2)}\rangle=\frac{1}{\sqrt{M^{(2)}}}(|\sqrt{\eta}\alpha,\sqrt{\eta}\alpha\rangle_{_{1,2}}|\sqrt{1-\eta}\alpha,\sqrt{1-\eta}\alpha\rangle_{_E}+\mu|\sqrt{\eta}\beta,\sqrt{\eta}\beta\rangle_{_{1,2}}|\sqrt{1-\eta}\beta,\sqrt{1-\eta}\beta\rangle_{_E}).
\ee
In order to study the noise effect on Wigner function we should trace out the mode 2 and environment mode $E$ by partial trace, i.e. $\rho_{1}^{(2)}=Tr_{2E}(|\Psi^{(2)}\rangle\langle\Psi^{(2)}|)$, then reduced density matrix in coherent basis $|\alpha\rangle$ and $|\beta\rangle$ reads
\be
\rho_1=\frac{1}{M^{(2)}}\{|\sqrt{\eta}\alpha\rangle\langle\sqrt{\eta}\alpha|+\mu^2|\sqrt{\eta}\beta\rangle\langle\sqrt{\eta}\beta|+\mu p^{2-\eta}(|\sqrt{\eta}\alpha\rangle\langle\sqrt{\eta}\beta|+|\sqrt{\eta}\beta\rangle\langle\sqrt{\eta}\alpha|)\}.
\ee
From Eq.(\ref{wigner}) we find the Wigner function for reduced density matrix of qubit like ECS after traveling through a noisy channel as
\be
W^{(2)}(\gamma)=\frac{2}{\pi M^{(2)}}\{e^{-2|\gamma-\sqrt{\eta}\alpha|^2}+\mu^2e^{-2|\gamma-\sqrt{\eta}\beta|^2}+\mu p^{2-\eta}
e^{-\frac{\eta}{2}(\alpha+\beta)^2}e^{-2|\gamma|^2}(e^{2\sqrt{\eta}(\gamma^*\alpha+\beta\gamma)}+e^{2\sqrt{\eta}(\gamma^*\beta+\alpha\gamma)})\}.
\ee
Setting $\lambda=x$ diagram of $W^{(2)}(x)$ as a function of $x$ for given $\eta,\alpha,\beta$ and $\mu=1$ is represented in figure (\ref{NW}b).
In order to study the noise effect on entanglement between modes $1$ and $2$, we should trace out the environment mode $E$ by partial trace, i.e. $\rho_{12}^{(2)}=Tr_{E}(|\Psi'^{(2)}\rangle\langle\Psi'^{(2)}|)$. By assumption $|\alpha\rangle$ and $|\beta\rangle$ are linearly independent the reduced density matrix in orthogonal basis $|0\rangle$ and $|1\rangle$ reads
\be
\rho_{12}^{(2)}=\frac{1}{2+2p^2}\left(
  \begin{array}{cccc}
    a_{11} & a_{12} & a_{12} & a_{14} \\
    a_{12} & a_{22} & a_{22} & a_{24} \\
    a_{12} & a_{22} & a_{22} & a_{24} \\
    a_{14} & a_{24} & a_{24} & a_{44} \\
  \end{array}
\right),
\ee
where
\be\begin{array}{l}
a_{11}=1+2p^2+p^{4\eta},\\
a_{12}=p^{-\eta} \sqrt{1-p^{2\eta}}(p^2 + p^{4 \eta}),\\
a_{14}=p^{2-2\eta}+p^{2\eta}-p^{4\eta}-p^2,\\
a_{22}=p^{2\eta}-p^{4\eta},\\
a_{24}=p^{\eta} (1-p^{2\eta})^{3/2},\\
a_{44}=(1-p^{2\eta})^2,
\end{array}
\ee
in which $p=e^{-\frac{1}{2}(\alpha-\beta)^2}$. Clearly this state is a two qubit mixed state and one of the suitable measure to evaluate the amount of entanglement is concurrence. For any two-qubit mixed state, concurrence is defined as $C=\max\{0,\lambda_{1}-\lambda_{2}-\lambda_{3}-\lambda_{4}\}$ where the $\lambda_{i}$'s are the non-negative eigenvalues, in decreasing order, of the Hermitian matrix $R=\sqrt{\sqrt{\rho}\tilde{\rho}\sqrt{\rho}}$,
with $\tilde{\rho}=(\sigma_{y}\otimes\sigma_{y})\rho^{*}(\sigma_{y}\otimes\sigma_{y})$
in which $\rho^{*}$ is the complex conjugate of $\rho$ when it is expressed in a standard basis and $\sigma_{y}$ represents the usual second Pauli matrix in a local basis $\{|0\rangle, |1\rangle\}$ \cite{Wootters1}. The concurrence is found as
\be
C^{(2)}(\rho)=\frac{p^2(-1+p^{-2\eta})}{1+p^2}.
\ee
Figure (\ref{NW}a) demonstrate the behaviour of concurrence for $\rho_{12}^{(2)}$ as a function of $\eta$. Figure (\ref{NW}a) shows that by decreasing noise parameter $\eta$ the concurrence of state $|\Psi^{(2)}\rangle$ is decreased. Figure (\ref{NW}b) yields the same result, i.e. by decreasing the $\eta$ two peaks close together and finally in $\eta=0$ we have one peak, which in turn implies that the state reduces to separable state as one would expect for complete decoherence.
\section{Conclusion}
In summary, we introduced two methods for generation of two-mode qubit and qutrit like ECS's. We  found that the one-mode Wigner functions for these ECS's reveal some information on the entanglement between modes in ECS's.  For qubit like ECS, it was shown that if $\alpha=\beta$ the state is separable and there is one peak in Wigner function $W^{(2)}(x,y)$. For $\alpha\neq\beta$, by increasing $\beta$ the distance of the peaks is also increased.
The same results arise  when we use concurrence measure. It was shown that  concurrence is a monotone function of separation of two peaks, $\Delta$, in which for $\Delta$  large enough concurrence tends to its maximum value ($C_{max}^{(2)}=1$) while for small values ($\Delta\rightarrow0$) the concurrence tends to zero. A similar result was discussed for qutrit like ECS, $|\Psi^{(3)}\rangle$. On the other hand we showed that the one-mode Wigner function $W^{(3)}(x,y)$ is entanglement sensitive. Finally we investigated the noise effects on the two-mode qubit like ECS. We recognized that by decreasing noise parameter $\eta$, the concurrence of state $|\Psi^{(2)}\rangle$ is decreased and at the same time two peaks on the Wigner function $W^{(2)}(x)$ approach each other. Ultimately for $\eta=0$ (maximum noise) the state is separable and one peak appears in the profile of Wigner function.

\par
\textbf{Acknowledgments}\\
The authors also acknowledge the support from the University of Mohaghegh Ardabili.

\end{document}